\documentclass[conference]{IEEEtran}
\IEEEoverridecommandlockouts

\usepackage{cite}
\usepackage{amsmath,amssymb,amsfonts}
\usepackage{graphicx}
\usepackage{textcomp}
\usepackage{xcolor}
\usepackage{hyperref}
\def\BibTeX{{\rm B\kern-.05em{\sc i\kern-.025em b}\kern-.08em
    T\kern-.1667em\lower.7ex\hbox{E}\kern-.125emX}}

\begin{document}

\title{Enhancing Student Feedback Using Predictive Models in Visual Literacy Courses
\thanks{We would like to thank the National Science Foundation (NSF) under the Division of Undergraduate Education (DUE) and the Directorate for Education and Human Resources (EHR) for their support through Grant No. 2216227.}
}

\author{\IEEEauthorblockN{Alon Friedman} 
    \IEEEauthorblockA{\textit{School of Information}\\ \textit{University of South Florida}\\
    Tampa, FL, USA\\ alonfriedman@usf.edu} \and 
    \IEEEauthorblockN{Kevin Hawley} 
    \IEEEauthorblockA{\textit{The Zimmerman School of Advertising}\\ \textit{and Mass Communications}\\ \textit{University of South Florida}\ Tampa, FL, USA\\ kevinhawley@usf.edu} \and 
    \IEEEauthorblockN{Paul Rosen} 
    \IEEEauthorblockA{\textit{Scientific Computing and Imaging Institute}\\ \textit{Kahlert School of Computing}\\ \textit{University of Utah}\\
    Salt Lake City, UT, USA\\ paul.rosen@utah.edu} \and 
    \IEEEauthorblockN{Md Dilshadur Rahman} 
    \IEEEauthorblockA{\textit{Scientific Computing and Imaging Institute}\\ \textit{Kahlert School of Computing}\\ \textit{University of Utah}\\
    Salt Lake City, UT, USA\\ dilshadur@sci.utah.edu}} 

\maketitle

\begin{abstract}
Peer review is a popular feedback mechanism in higher education that actively engages students and provides researchers with a means to assess student engagement. However, there is little empirical support for the durability of peer review, particularly when using data predictive modeling to analyze student comments. This study uses Na\"ive Bayes modeling to analyze peer review data obtained from an undergraduate visual literacy course over five years. We expand on the research of Friedman and Rosen ~\cite{friedman2021leveraging} and Beasley et al. ~\cite{beasley2021through} by focusing on the Na\"ive Bayes model of students' remarks.
Our findings highlight the utility of Na\"ive Bayes modeling, particularly in the analysis of student comments based on parts of speech, where nouns emerged as the prominent category. Additionally, when examining students' comments using the visual peer review rubric, the lie factor emerged as the predominant factor. Comparing Na\"ive Bayes model to Beasley's approach ~\cite{beasley2020sentiment}, we found both help instructors map directions taken in the class, but the Na\"ive Bayes model provides a more specific outline for forecasting with a more detailed framework for identifying core topics within the course, enhancing the forecasting of educational directions. Through the application of the Holdout Method and $\mathrm{k}$-fold cross-validation with continuity correction, we have validated the model's predictive accuracy, underscoring its effectiveness in offering deep insights into peer review mechanisms. Our study findings suggest that using predictive modeling to assess student comments can provide a new way to better serve the students' classroom comments on their visual peer work. This can benefit courses by inspiring changes to course content, reinforcement of course content, modification of projects, or modifications to the rubric itself.
\end{abstract}


The use of graphics is a growing trend in textbooks, scholarly journals, and social media. In response to the growing demand for data visualization skills, higher education has begun offering courses in data visualization~\cite{Chegu20116communication}. A more recent development in visualization education is the use of visual peer review, which allows students to provide feedback on each other's data graphics work. Analyzing student comments on their peers' work, however, can be challenging. With the continuous advancement of technology facilitating faster data analysis and introducing new data modeling approaches, researchers are increasingly turning to natural language processing (NLP) as the preferred platform to gain deeper insights into student comments. 

Peer review is a valuable technique for enhancing composition writing by providing critical feedback. In education, it finds widespread application in liberal arts courses and is employed in professional settings, such as expert code reviews and academic publications. In the context of visualization education, peer review facilitates collaborative learning among students, allowing them to receive feedback from their peers. Furthermore, this evaluative process presents an opportunity for students to strengthen their understanding of recently acquired course concepts. 

Friedman and Rosen~\cite{friedman2021leveraging} developed a comprehensive rubric for visual peer review, focusing on assessing the effectiveness of visualizations and facilitating student skills development. The rubric was designed to enhance the precision of visual representations through systematic evaluation and feedback. Students are instructed to use this rubric for critical evaluation and constructive feedback on their peers' work. To obtain quantitative scoring metrics from the visual peer review rubric, including the overall assessment of natural language processing (NLP) in the reviews (i.e., positive or negative sentiment), parts-of-speech counts, and average comment length, Beasley et al.~\cite{beasley2021through} conducted an evaluation. They employed sentiment analysis techniques to generate positive and negative sentiment-bearing keywords, which were then algorithmically matched to analyze the data. This model, however, was not designed to predict students' comments. 

In this study, we used Na\"{i}ve Bayes modeling to address the need for prediction modeling in visualization education and the evaluation of student feedback. Our approach aimed to bridge existing research gaps by demonstrating the predictive potential of Na\"{i}ve Bayes modeling, especially when applied to student feedback in the context of visualization education. This modeling can then be used to improve courses by inspiring changes to course content, reinforcing existing material, modifying projects, or even refining the rubric itself. We particularly sought to address the following hypotheses:

\begin{itemize}
    \item \textbf{H1:} Na\"{i}ve Bayes classification model can accurately predict the parts of speech students used with an accuracy of at least 85\% and above. 

    \item \textbf{H2:} The accuracy of Na\"{i}ve Bayes in predicting student visual peer review rubric will be 85\% or higher.

    \item \textbf{H3:} The validity of Na\"{i}ve Bayes classification for both cases (part of speech and visual peer rubric) employing the Holdout Method and $\mathrm{k}$-fold cross-validation will be judged by its accuracy of 85\% or above.

\end{itemize}

In the pursuit of advancing our understanding of peer review mechanisms within visualization education, this study is anchored in the analysis of five years of peer review data. In setting our hypotheses, we aimed for an accuracy threshold of at least 85\% for the Naïve Bayes classification model. This benchmark, selected based on work by Sorour et al. \cite{sorour2014predicting} and the inherent complexities of educational data, represents a rigorous yet attainable goal that acknowledges the challenges of accurately predicting student feedback dynamics. 

The exploration of predictive modeling within this context not only enriches our understanding of student engagement in peer review processes but also contributes to the broader discourse on visual communication education. Initially, we delve into the foundations of visual communication education and its intersection with peer review, setting the stage for a detailed examination of Natural Language Processing (NLP) applications in analyzing student feedback. Subsequently, the paper provides an overview of our course content, offering a comprehensive utilizing predictive modeling venture. To improve peer review in education, we aim to illuminate pathways that can be used to improve the effectiveness of future classes and pedagogical strategies through this study.

\section{Visual Communication Education}
In 2006, the ACM Siggraph Education Committee discussed the purpose of visualization in education. In their summary of the topic, Domik~\cite{domik2009my} posted the question: ``Do we need formal education in visualization?" She answered this question with an affirmative, ``yes,'' emphasizing the importance of formal education in visualization. However, in her discussion, Domik~\cite{domik2009my} did not address the crucial aspect of student assessment within the classroom. In higher education research, student assessment has emerged as a fundamental component for measuring the success of effective learning outcomes. This emphasis on assessment has also led to the adoption of innovative data modeling techniques to analyze and evaluate educational data~\cite{okoye2022towards}. The current surge in online education has further emphasized the need to actively engage students in virtual classrooms to ensure meaningful and beneficiary learning experiences.

During the examination of student engagement in visualization education, it was observed that students exhibited higher levels of engagement when visualization techniques were applied~\cite{trogden2023mapping}. The increase in online education has amplified the importance of engaging students effectively in virtual classrooms while also providing mechanisms to measure their productivity. Furthermore, researchers identified two main themes in visualization education. The first is the content of the course. According to Firat et al.~\cite{firat2022interactive}, the field encompasses the use of visual elements, design principles, and storytelling techniques to communicate complex information, ideas, and concepts visually, making them more accessible, understandable, and engaging to a wide audience. The second theme is the subjective evaluation of the quality and accuracy of visualizations, specifically, training students to critically evaluate others' visualizations, which is the focus of our work. These essential skills are often taught through informal methods, such as self-questionnaires and group discussions. However, traditional evaluation methods often lack active participation, which may hinder students' development of essential skills such as assessing and providing feedback to others, as well as cultivating the ability to self-assess and enhance their own work~\cite{liu2020information}. The significance of ``collaborative visualization''~\cite{isenberg2011collaborative}, which entails contributions from several experts toward a common aim of comprehending the visual object, phenomena, or data under inquiry, does not address student engagement in the class. Friedman and Rosen~\cite{friedman2021leveraging} introduced visual peer review in visualization education in response to student peer review's rising popularity in higher education. Beasley ~\cite{beasley2020sentiment} developed a sentiment analysis model based on student peer review examined through visual peer review ~\cite{beasley2021through}.  
early

\section{NLP Modeling to Examine Peer Review for Students}
Recently, in Artificial Intelligence (AI), supervised machine learning models have been applied to various prediction scenarios in education research, including predicting student engagement and student grades. These models are trained on data that include student characteristics, such as prior academic performance, demographic information, and course-related variables~\cite{chen2022two}. Kahn et al.~\cite{khan2021artificial} found that using prediction algorithms to track student performance in the classroom can help to identify low-performing students early in the learning process. Those approaches analyze students' comments typically by using rules that match the feedback to a set of predefined linguistic patterns, as well as parts-of-speech (POS) tagging and a carefully crafted thesaurus relevant to each domain. 

Zingle et al.~\cite{zingle2019detecting} reported on several of these methodologies that can be applied to student peer review under an NLP framework. They included in their review: rule-based NLP methods, support vector machines, Na\"ive Bayes, and many others. Rule-based NLP methods use POS tagging to determine the word class of each processed word. The relevant tags used for this classifier include nouns, verbs, adjectives, adverbs, and so on. The drawback of those NLP methods is that the result remains biased, according to Li et al.~\cite{li2021robustness}.
The next model they reported on was support vector machines, known for their accuracy and ability to handle large datasets. However, according to Xiao et al.~\cite{xiao2020detecting}, they are not sensitive to outliers and can be hard to interpret. The third model they covered was the Na\"ive Bayes. It is a probabilistic model that assumes that the presence of a particular feature in a document is independent of the presence of any other feature. This assumption of independence makes the model relatively simple to train and interpret. It also provides a path for a new model by classifying the data using a train-and-test split of 90\% training data and 10\% testing data. Transformers are used to facilitate the construction of the classifier. These transformers include a count vectorizer that converts a collection of text documents to a matrix of token counts for future use. Brodesen et al.~\cite{brodersen2015inferring} reported on the success of using Bayesian modeling in time series by examining econometrics markets and its implementation of the model for prediction.

In our early studies~\cite{beasley2021through}, we used~\cite{beasley2020sentiment} se model to analyze students' comments. This model allows us to match positive and negative sentiment-bearing keywords to create metrics, including counts of parts of speech and average comment length, as well as the overall sentiment of the review (positive or negative). Using aspect extractors was part of the algorithm, which scanned text in a sliding window and produced a list of significant aspects (nouns) close to emotive terms (adjectives) in the text. Because of the intricacy of the analysis and our approach, Firat et al.~\cite{firat2022interactive} reported on the difficulty of evaluating visual peer review data using our model. In this study, we concentrated on the Na\"ive Bayes model, which many educational researchers use for their examinations~\cite{razaque2017using, wahyuni2020implementation, tripathi2019naive}. More specifically, Maitra et al.~\cite{maitra2018mining} reported using this model to analyze student feedback. Our study used this model because of its simplicity and ability to predict and verify results.

\section{Courses Overview and Content}

The University of South Florida began building a visual communication curriculum across the College of Arts and Sciences disciplines in the Fall of 2017, offered during the Fall and Spring semesters. The College of Arts and Sciences faculty focus on teaching visualization as a technique for creating images, diagrams, or animations to communicate a message using different platforms and programming languages. The primary objective of this survey course is to equip students with a fundamental understanding of the diverse manifestations of visual communication while offering practical opportunities to engage in project development across various domains. It covers a wide range of areas, including visual persuasion, data visualization, and visual storytelling.

The assignments for this class were based on projects that gave students hands-on experience in applying theoretical concepts to design and develop projects. In the data visualization module, we used a visual rubric that focused on: 1) clear labeling, 2) the lie factor, 3) data-to-ink ratio, 4) chart junk, and 5) Gestalt principles~\cite{todorovic2008gestalt}. The first category asked the students to search their peer’s visualization for clear and detailed labeling of every aspect of the data represented in the graph or chart. The next three categories were based on Tufte’s principles of design~\cite{tufte2006beautiful}. The lie factor directs the relationship between the size of the effect shown in the graphic and the size of the effect shown in the data. The data-ink ratio, the proportion of ink (or pixels) used to present data compared to the total ink used in the display, refers to the non-erasable ink used to present data. If data ink were removed from the image, the graphic would lose its content. The next category, known as chart junk, refers to all of the visual elements in charts and graphs that are not necessary to comprehend the information represented on the graph or that distract the viewer from this information~\cite{Tufte:1983}. The last category is Gestalt principles. Those principles describe how the human eye perceives visual elements, which tells us that complex images tend to be reduced to simpler shapes.

\begin{figure}[tb]
    \vspace{-5.5ex} 
    \centering
    \includegraphics[width=0.9\linewidth]{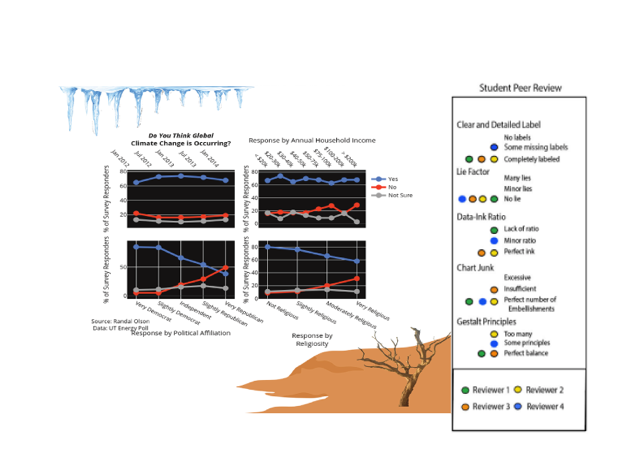}
    \caption{Student map-making project and visual peer review rubric the student evaluate the work.}
    \label{fig2}
\end{figure}

Students submitted their assignments and engaged in peer review using a specific form, with the grades contributing 10\% towards their overall grade. They used Google Fusion Tables and Adobe Illustrator to create a map-making project, which taught them to integrate raster and vector graphics. This project involved applying vector graphics such as words, lines, and arrows to add context to the image content. \autoref{fig2} presents an example of a student project, highlighting their interest and active involvement in the subject, despite its imperfections and illustrating their engagement with the rubric provided for reviewing the work.

\begin{figure}[tb]
    \centering
\includegraphics[width=0.9\linewidth]{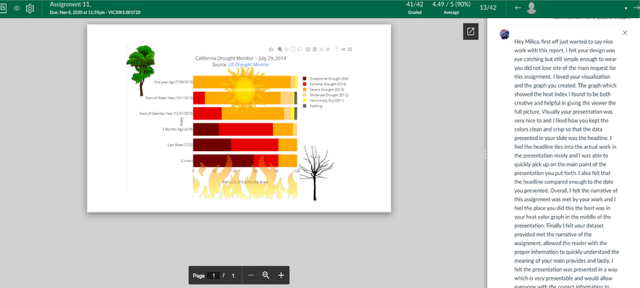}
    \caption{Example of a student's comments on their peer's work.}
    \label{fig3}
\end{figure}

\section{Study Methodology}

Our study investigated the role of student comments in visual peer review, with a specific emphasis on the utilization of parts of speech as used in their visual peer review rubric comments. Our methodology for the efficacy of student comments within visual peer review processes leverages was based on a statistical modeling methodology as outlined by \cite{dvir2023informal}. 
In our earlier research, we found that students highly valued peer reviews and actively participated in the feedback process, in comparison to their relatively lower emphasis on self-reflection or seeking input from others~\cite{beasley2021through}. The algorithm we used from our previous studies matched the sentiment-bearing part of speech to produce metrics that included the overall sentiment of the review counts of parts of speech (i.e., noun, adjective, adverb, etc.) and the average length of comments.

At the current study, we have identified the Na\"{i}ve Bayes modeling technique as particularly suited to address the complexities of analyzing textual data within educational feedback. This choice was motivated by the algorithm's proven capability to efficiently handle and interpret the varied and nuanced data represented by student comments. Recognizing the challenges inherent in predicting educational outcomes from textual feedback, we set an accuracy goal of 85\% for our model. This target was established based on a comprehensive review of related literature and a detailed assessment of our dataset, balancing ambition with the practicalities of our unique analytical context.

\section{Study Procedure}
In order to evaluate the effectiveness of the Naive Bayes model, this study utilized both training and test sets. The training set was used to train the model, while the test set was used to evaluate the model’s predictions on new, unseen data. This evaluation is critical for gauging the model’s overall effectiveness and its ability to generalize beyond the training data.
The study followed procedure of assumable the 
Na\"{i}ve Bayes modeling. 
 We first trained a Na\"{ı}ve Bayes classifier model on a corpus of text that had been labeled with its sentiment (i.e., numeric values from -1 to 1, representing negative, neutral, or positive). The model examines the probability of each word appearing in a particular part of speech in the student’s comments. Once the model was trained, we used it to predict the sentiment of new text.

By employing the Na\"{i}ve modeling technique, we aimed to harness its strengths in processing and analyzing textual data derived from student feedback. This equation represents the core of our predictive modeling, illustrating how the probability of each outcome. This equation represents the core of our predictive modeling, illustrating how the probability of each outcome (\(Y\)) is calculated based on the conditional probabilities of various input variables (\(X_1\) to \(X_n\)) using the Na\"{i}ve Bayes equation as follows:

\noindent
\resizebox{\linewidth}{!}{
\begin{minipage}{\linewidth}
\begin{align*}
P(Y&=k | X_1, X_2, \ldots, X_n)=\\ 
&\frac{P(X_1 | Y=k) \cdot P(X_2 | Y=k) \cdot \ldots \cdot P(X_n | Y=k) \cdot P(Y=k)}{P(X_1) \cdot P(X_2) \cdot \ldots \cdot P(X_n)}
\end{align*}
\end{minipage}
}
\hfill \break

After establishing the theoretical framework, the study employed the Na\"{i}ve Bayes modeling technique, utilizing the \texttt{e1071} package in R for generating algorithms capable of assessing probabilities of parts of speech within the visual peer review rubric based on predictor variables \cite{meyer2019package}.

The following step in our study delves into two pivotal sections: Model Validation and Evaluation Metrics. These sections are integral to our research as they outline the rigorous methods employed to assess the effectiveness and reliability of the Na\"{i}ve model applied within our framework.

\subsection{Model Validation}

The model's validation was performed through the Holdout Method and \(\mathrm{k}\)-fold Cross-validation to ensure its robustness and reliability. The Holdout Method involved dividing the dataset into a training set, comprising 70-80\% of the data for model training, and a testing set, making up the remaining 20-30\%, for evaluating the model's performance on unseen data. Additionally, \(\mathrm{k}\)-fold Cross-validation was utilized to rigorously assess the model's performance across different data segments, enhancing the reliability of our findings \cite{dake2023using}.

\subsection{Evaluation Metrics}
The effectiveness of the Na\"{i}ve Bayes model was meticulously evaluated through its performance on the testing set, with particular attention to Sensitivity, Specificity, Positive Predictive Value (PPV), and Negative Predictive Value (NPV). These metrics used in determining the model's precision under constructive or non-constructive feedback. Sensitivity and Specificity provided insights into the model’s ability to identify relevant feedback accurately, while PPV and NPV assessed the reliability of these classifications. 
A confusion matrix, generated using the \texttt{confusionMatrix} function from the \texttt{caret} package in R, facilitated the calculation of Sensitivity, Specificity, Positive Predictive Value (PPV), and Negative Predictive Value (NPV). These metrics provided a comprehensive overview of the model's performance, enabling an evaluation of its accuracy in classifying feedback as either relevant or non-relevant.

The metrics derived from the confusion matrix were instrumental in quantifying the model's ability to accurately identify and classify the student feedback within the testing set. This analysis was critical for gauging the effectiveness of the Na\"{i}ve Bayes model in parsing constructive from non-constructive feedback, highlighting its potential to offer meaningful insights into educational feedback mechanisms.

\section{Data Collection}
The study examined the part of speech and peer review score with respect to student comments on the rubric. To illustrate this scenario, \autoref{fig3} shows the capture of a student's work and their peer's review comments.

\begin{figure}[tb]
    \centering
    \vspace{-5.5ex} 
    \includegraphics[width=0.80\linewidth]{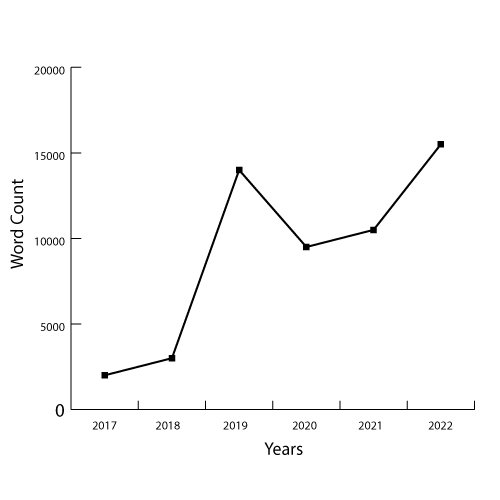}
    \caption{Summary of the word counts from 2017 to 2022.}
    \label{fig4}
\end{figure}

A total of 3082 students participate in this study. The total word count of all the comments students submitted was 655,816, with an average of 21 words per review.  Additionally, we conducted an analysis of student word count trends over time. Interestingly, we observed two significant peaks during the five-year data collection period: one in 2019 and another in 2022. We attribute the surge in word count during 2019 to the emergence of the COVID-19 pandemic, which likely influenced the nature and length of peer review comments. In 2022, we experienced the same growth in the number of words used. \autoref{fig4} summarizes word count from 2017 to 2022.


\begin{figure}[tb]
    \centering
    \includegraphics[width=0.80\linewidth]{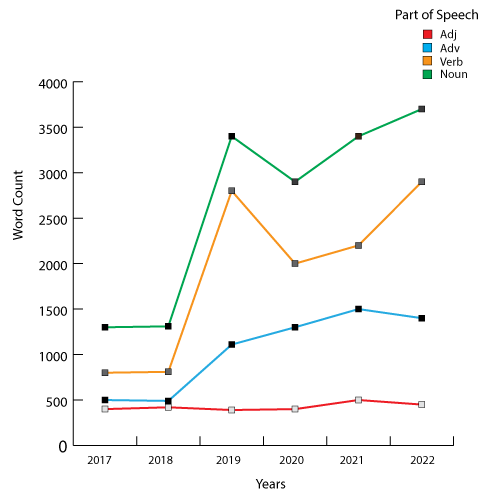}
    \caption{Summary of parts of speech from students' comments.}
    \label{fig5}
\end{figure}

We found the same patterns when we analyzed students' peer review part of speech. Specifically, our analysis consistently revealed that nouns were the most frequently occurring part of speech among students' peer review comments. This trend persisted consistently throughout the entire duration of our examination, spanning multiple years. \autoref{fig5} summarizes our findings. 

\section{Results}

The first step in our analysis was to train our Na\"ive Bayes classification model, which allowed us to create conditional probabilities for all the variables in our dataset. By using a subset of our dataset to create these conditional probabilities, we were able to calculate probabilities that accurately represented the distribution of our data. Our results showed that the Na\"ive Bayes model accurately predicted parts of speech and visual peer review rubric, achieving an accuracy of at least 90\% on both the training and test sets.
For our study, the Training Set encompassed students' comments, part of speech, and visual peer review rubric score. The Na\"ive Bayes model successfully identified the primary parts of speech and key components of our rubric. Furthermore, the model facilitated the customization of individual students' scores in both classifications, thereby providing a more precise and tailored evaluation of their performance. Overall, our findings demonstrate the efficacy of the Na\"ive Bayes classification model as a valuable tool for analyzing student data and enhancing assessment outcomes. By accurately predicting parts of speech and visual peer review rubric, the model enables the identification of areas of strength and weakness in student work, as well as the personalization of instruction to better cater to individual student needs.


\subsection{Parts of Speech and Na\"ive Bayes Classification}
Under \textbf{H1}, we measured the accuracy of the Na\"ive Bayes classification model in predicting parts of speech (X), achieving an accuracy of at least 75\%. However, our study revealed that the model achieved a much higher accuracy of 99\% with a very low p-value, indicating strong correlation between the predicted parts of speech (X) and the actual observed parts of speech (Y). Therefore, we accept our hypothesis that the Naive Bayes model accurately predicts the part of speech used by students.



\begin{figure}[tb]
    \centering
    \includegraphics[width=0.9\linewidth]{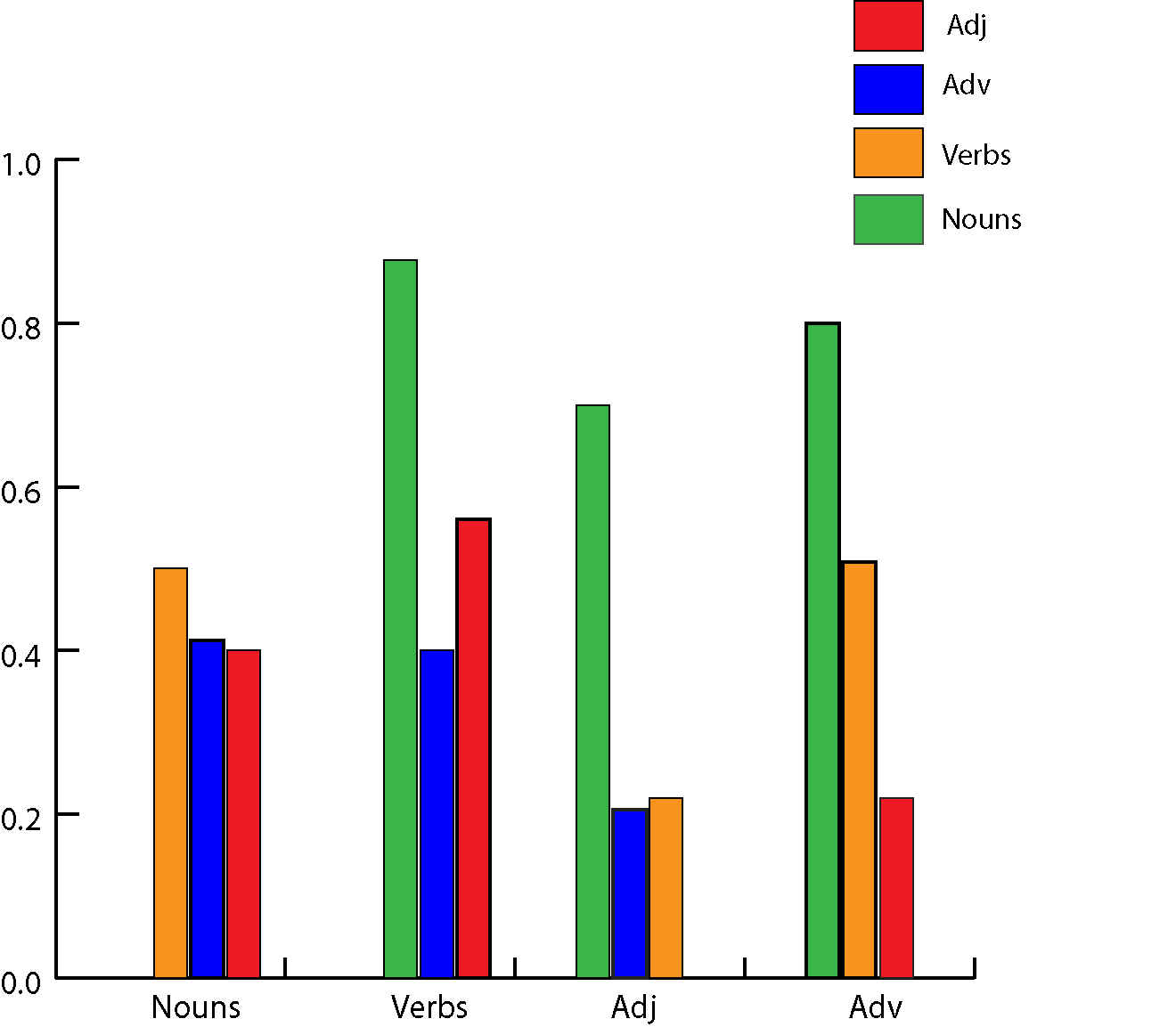}
    \caption{Summary of the correlation matrix score on parts of speech (nouns, verbs, adverb (adv), and adjective (adj)).}
    \label{fig6}
\end{figure}

Using this model, we were able to extend our analysis and report on the most common parts of speech that students use in their comments. We found that nouns were the most common part of speech, with students using two nouns per sentence on average. The length of the comments was typically five words. \autoref{fig6} summarizes the correlation matrix score for our findings.

\subsection{Analyzing Peer Review Rubric and Na\"ive Bayes Classification}

Under \textbf{H2}, we investigated the impact of incorporating the visual peer review rubric on the accuracy of the Na\"ive Bayes model in predicting student grades (Y). Our analysis revealed that incorporating the visual peer review rubric significantly improved the accuracy of the Na"ive Bayes model by 85\% with respect to the predicted student visual peer reviews (X). Under Na"ive Bayes model achieved a high accuracy of at least 90\% on both the training and test sets. In this study, the training set represents a subset of the data used to train the Na\"ive Bayes model, and the test set is a separate subset of data used to evaluate its performance.

\begin{figure}[tb]
    \centering
    \includegraphics[width=0.9\linewidth]{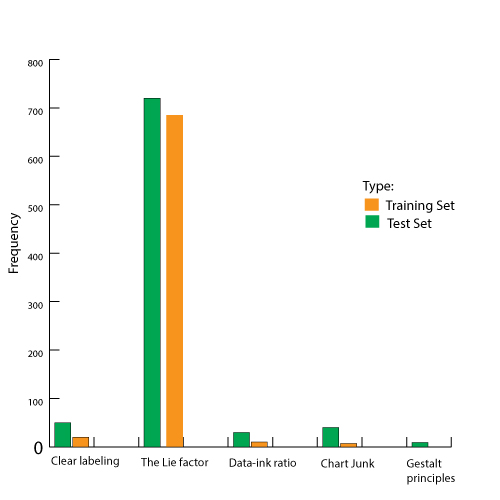}
    \caption{The Na\"ive Bayes predictive model on visual peer review rubric}
    \label{fig:teaser}
\end{figure}

Moreover, the model successfully predicted all five categories of the rubric, with the lie factor demonstrating a strong correlation with the part of speech noun score. The model was also able to predict all five categories of the rubric, but the score for the lie factor stood out as being highly correlated with the noun part of speech score. \autoref{fig:teaser} shows the confusion matrix for the Na\"ive Bayes model for the peer review rubric. We found that the lie factor had the highest score of 0.99, compared to the other four categories. The results were consistent across both datasets, the training and test sets. 
Therefore, we accept our second hypothesis that holds the Naive Bayes model accurately predicts the accuracy of the visual peer review rubric. These findings indicate that utilizing the visual peer review rubric can assist instructors in identifying students' usage of part of speech and enhance their ability to convey meaning in their communication with peers and instructors ~\cite{Rahmah:2018}. Furthermore, we recommend that instructors pay close attention to the lie factor, as it is also significant to students.

\subsection{Validation of the Study}

Under \textbf{H3}, the validation of our Na\"ive Bayes predictive model's accuracy was meticulously conducted using the Holdout Method alongside $\mathrm{k}$-fold cross-validation with continuity correction. This dual approach allowed us to rigorously compare two proportions of the model's predictions against actual outcomes, focusing on both parts of speech analysis and the visual peer review rubric.

Our analysis yielded strong performance metrics, indicating the model's effectiveness. It achieved a Sensitivity rate of 88\% and a Specificity rate of 87\%, showcasing its capability to accurately classify relevant and non-relevant comments. This underscores the model's high precision in parsing student feedback.

However, the Positive Predictive Value (PPV) and Negative Predictive Value (NPV) presented challenges, with scores of 30\% and 23\%, respectively. These lower values suggest limitations in the model's predictive precision and reliability, particularly in its capacity to accurately classify comments as positive or negative instances based on the predefined criteria.

Despite these challenges, the overall accuracy of the Na\"ive  model, as determined through $\mathrm{k}$-fold cross-validation, was impressive, showing an accuracy rate of 95\% with a standard deviation of 2\%. This high level of accuracy, combined with the stability of correlation coefficients ranging from 0.82 to 0.87 in sensitivity analysis, confirms the model's robustness and its consistent performance across various testing scenarios.

The observed correlation, stable across different parameter adjustments, reaffirms the model's reliability in capturing the nuanced relationship between the accuracy of the visual peer review rubric (variable X) and the Na\"ive Bayes model predictions (variable Y). Such findings are pivotal, as they not only validate the hypotheses underpinning this study but also demonstrate the predictive model's utility in enhancing educational assessment and feedback strategies. As summarized in Table \ref{tab:model_performance}, the Na\"ive Bayes model exhibited strong Sensitivity and Specificity, indicating its effectiveness in classifying comments accurately. However, the lower PPV and NPV scores highlight potential areas for refining the model to enhance its predictive precision 

\begin{table}[tb]
\centering
\caption{Summary of Na\"ive Bayes Model Performance Metrics}
\resizebox{\linewidth}{!}{
\begin{tabular}{lcl}
\hline
\textbf{Metric} & \textbf{Value (\%)} & \textbf{Comments} \\
\hline
Sensitivity & 88 & \parbox[t]{2.75cm}{\raggedright High ability to detect relevant comments\strut} \\
Specificity & 87 & \parbox[t]{2.75cm}{\raggedright  High ability to recognize non-relevant comments\strut} \\
Positive Predictive Value (PPV) & 30 & \parbox[t]{2.75cm}{\raggedright Indicates room for improvement\strut} \\
Negative Predictive Value (NPV) & 23 & \parbox[t]{2.75cm}{\raggedright Indicates room for improvement\strut} \\
Overall Accuracy & 95 & \parbox[t]{2.75cm}{\raggedright High accuracy in predictions\strut} \\
Standard Deviation (Accuracy) & 2 & \parbox[t]{2.75cm}{\raggedright Indicates model stability\strut} \\
\hline
\end{tabular}
}
\label{tab:model_performance}
\end{table}

The implications of these findings are significant for visualization instructors and educational practitioners. By leveraging the Na\"ive Bayes model, educators can gain deeper insights into student engagement and the effectiveness of peer reviews, facilitating a more personalized and nuanced approach to teaching and learning. Moreover, the analysis of peer review data through this predictive lens offers a pathway to identifying long-term trends and areas for improvement within educational programs.

The result of PPV and NPV scores highlight areas for further model refinement, the overall validation process underscores the Na\"ive Bayes model's value as a sophisticated tool in the landscape of educational analytics. The model's ability to accurately predict parts of speech and assess peer review rubrics presents a novel avenue for advancing pedagogical practices and fostering student success.

\section{Limitations and Risks} 
The findings of the study offer compelling evidence supporting the notion that Na\"{i}ve Bayes model possesses the capability to accurately predict data within the context of parts of speech and visual peer review rubrics. However, the study did not consider instructors' input or their justification for the peer review rubric. As we proceed with the integration of this framework within our own classroom, we remain attentive to a few potential issues. Primarily, student peer review is still a supplement to established educational methodologies. Consequently, the instructor's active involvement is crucial in both incorporating this system into their classroom and analyzing student comments to acquire novel insights into their students.
More specific challenges that we are anticipating: 
1) Instructor workload: The instructor will need to take on more work to implement this system. They will need to create and maintain the visual peer review rubric, as well as analyze student comments.
2) Bias: There is a risk that the visual peer review rubric could be biased. For example, it could exhibit bias towards certain types of data visualizations or favor wordy rubric descriptions.
As artificial intelligence (AI) continues to gain prominence in educational settings, it will significantly influence the teaching of data visualization and the analysis of student work. Therefore, our objective is to develop and implement a visual peer review system that not only aligns with the instructor's workload but also effectively detects and addresses biases. The integration of AI technologies can enhance the system's capabilities in these aspects, but it is important to establish a solid foundation and understanding of the basic principles before incorporating advanced AI techniques.

\section{Conclusion}

We used a predictive data modeling approach to evaluate student peer review in visualization courses. We used Na\"ive Bayes modeling to analyze students' speech components and rubric scores. The data model we used was trained on binary data collected from our Introduction to Visual Literacy class over a five-year period. This approach not only assesses the rubric's effectiveness in clarifying students' review comments, but it also addresses the challenges of larger class sizes by enabling comprehensive analysis of student feedback and peer evaluations.

Our evaluation of the Na\"ive Bayes model on both training and test sets confirmed the robustness of our findings and the validity of our research approach. In our study, all three hypotheses were supported, indicating that the model is accurate at predicting the accuracy of the visual peer review rubric and student review comments. This finding has important implications for visualization instructors, as it provides a reliable tool for assessing student learning.

As AI gains prominence in visual education and student data analysis, our use of the predictive model can enhance our understanding and interpretation of student data. By integrating visual peer review, instructors can access deeper insights, which can foster improved communication and enriched learning outcomes within the classroom. This approach can also inspire potential modifications to course content, projects, or rubrics, while also highlighting the value of Na\"ive Bayes modeling in our study.





\end{document}